\documentclass[usenatbib]{mnras}
\usepackage[T1]{fontenc}
\usepackage{ae,aecompl}
\usepackage{bm}
\usepackage{amsmath,amssymb}
\usepackage{dcolumn}
\usepackage[dvips]{graphicx}
\usepackage{epsf}
\usepackage{color}
\usepackage{epsf}
\usepackage{epsfig}
\usepackage{graphicx}
\usepackage{longtable}
\usepackage{psfrag}
\usepackage{txfonts}
\usepackage{newtxtext,newtxmath}

\newcommand{\be}{\begin{equation}}
\newcommand{\ee}{\end{equation}}
\newcommand{\ba}{\begin{eqnarray}}
\newcommand{\ea}{\end{eqnarray}}

\title[Cosmic acceleration with future surveys]{{Evidence for cosmic acceleration with next-generation surveys:\\ A model-independent approach}}

\author[C. A. P. Bengaly et al.]
{
 \parbox{\textwidth}{
  Carlos A. P. Bengaly$^{1}$\thanks{E-mail: \texttt{Carlos.Bengaly@unige.ch}}
 }
 \vspace{0.4cm}\\
 \parbox{\textwidth}{
 $^{1}$Department of Physics \& Astronomy, University of the Western Cape, Cape Town 7535, South Africa\\
 $^{2}$ D\'epartement
 de Physique Th\'eorique, Universit\'e de Gen\`eve, 24 quai Ernest Ansermet, 1211 Gen\'eve 4, Switzerland
 }
}

\begin{document}
\label{firstpage}
\pagerange{\pageref{firstpage}--\pageref{lastpage}}
\maketitle

\begin{abstract}
We quantify the evidence for cosmic acceleration using simulations of $H(z)$ measurements from SKA- and Euclid-like surveys. We perform a non-parametric reconstruction of the Hubble parameters and its derivative to obtain the deceleration parameter $q(z)$ using the Gaussian Processes method. This is a completely model-independent approach, so we can determine whether the Universe is undergoing accelerated expansion {\it regardless} of any assumption of a dark energy model. We find that Euclid-like and SKA-like band 1 surveys can probe cosmic acceleration at over $3$ and $5\sigma$ confidence level, respectively. By combining them with a SKA-like band 2 survey, which reaches lower redshift ranges, the evidence for a current accelerated phase increases to over $7\sigma$. This is a significant improvement from current $H(z)$ measurements from cosmic chronometers and galaxy redshift surveys, showing that these surveys can underpin cosmic acceleration in a model-independent way. 
\end{abstract}

\begin{keywords}
Cosmology: observations; Cosmology: theory; (cosmology:) large-scale structure of the Universe; 
\end{keywords}

\section{Introduction}\label{intro}

The evidence of late-time cosmic acceleration is one of the biggest scientific discoveries of the last decades~\citep{Riess:1998cb, Perlmutter:1998np}. It is ascribed to dark energy which accounts for roughly 68\% of the material content of the Universe~\citep{Aghanim:2018eyx}. The best candidate to explain this phenomenon is the so-called Cosmological Constant $\Lambda$, which is commonly associated with the vacuum density energy of the Universe. Combined with cold dark matter, responsible for cosmic structure formation, we have the $\Lambda$CDM model, i.e., the standard model of Cosmology at the present moment. Although the $\Lambda$CDM model is able to provide the best explanation for the cosmological observations thus far, it is plagued with coincidence and fine-tuning problems. Since its "discovery", many attempts have been envisaged to address these issues - see~\cite{Li:2011sd, Clifton:2011jh} for reviews on this topic. Nonetheless, $\Lambda$CDM remains the best candidate we have. 

Given the still unknown nature of the current accelerated expansion, it is essential to quantify how well we can detect this phenomenon, since it could rule out the standard model, and even the possibility of the existence of the dark energy paradigm as a whole. Within the context of the standard model, cosmological observations show that the Universe is currently accelerating at roughly $5\sigma$ level~\citep{Shapiro:2005nz,Ishida:2007ej,Giostri:2012ek,Santos:2015nua,Rubin:2016iqe,Haridasu:2017lma,Tutusaus:2017ibk,Lin:2017yfl,Rubin:2019ywt}, albeit this result was recently disputed using the Hubble diagram of Type ia Supernovae~\citep{Nielsen:2015pga,Ringermacher:2016gpa,Dam:2017xqs,Colin:2018ghy,Rameez:2019nrd,Colin:2019ulu} and quasars~\citep{Lusso:2019akb,Yang:2019vgk,Velten:2019vwo}. Other works looked at model-independent probes of cosmic acceleration, thus independent of dark energy, and even General Relativity assumptions, such as kinematic analyses~\citep{Rapetti:2006fv, Cunha:2008ja,Carvalho:2011qw,Lu:2011ue,Nair:2011tg,Muthukrishna:2016evq,Jesus:2017cyo,Capozziello:2017nbu,Heneka:2018tta}, besides non-parametric approaches~\citep{Mortsell:2008yu,Blake:2011,Velten:2017ire,Haridasu:2018gqm,Tutusaus:2018ulu,Gomez-Valent:2018gvm,Jesus:2019nnk,Arjona:2019fwb}. These works found at least moderate evidence ($> 2\sigma$) for present time accelerated expansion using existing data. 

In this work, we rely on the latter approach to probe the current cosmic acceleration evidence. We adopt a non-parametric method called Gaussian Process for this purpose. We focus on forecasting how well Hubble parameter measurements, $H(z)$, from future redshift surveys mimicking the specifications of Euclid galaxy and SKA intensity mapping - hereafter Euclid-like and SKA-like surveys, respectively. These measurements can be obtained from the Baryonic Acoustic Oscillation scale of the galaxy clustering (see~\citealt{Bengaly:2019oxx} for a thorough explanation about this). 

Rather than focusing on the {\it value} of $q_0$, as many authors do, we shall focus on determining the {\it uncertainty} of its measurement - and hence how well can we probe the evidence for a positive accelerated phase of the Universe today. To do so, we perform a non-parametric reconstruction of the deceleration parameter $q(z)$, which is a quantity that directly depends on $H(z)$ and its first derivative with respect to the redshift, in order to determine $q_0$, its current value. We find that these surveys will deliver significant improvement on $q_0$ constraints compared to current observational data, and that this result is totally independent of the fiducial dark energy model assumed.

\section{Data Analysis}\label{data}


\begin{figure*}
\centering
\includegraphics[width=0.48\textwidth,
height=5.8cm]{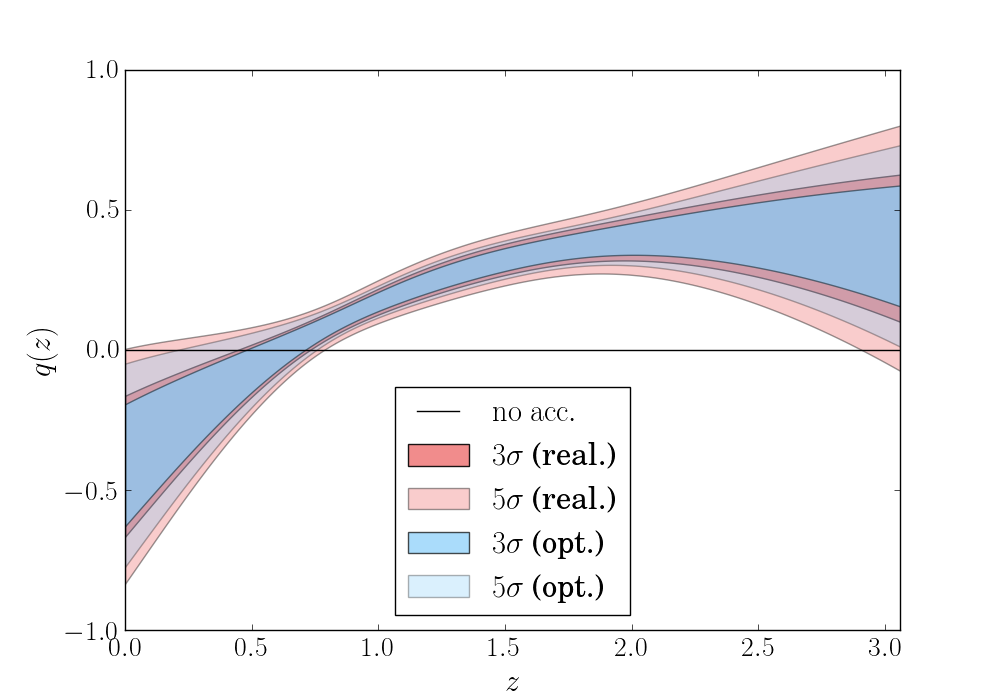}
\includegraphics[width=0.48\textwidth,
height=5.8cm]{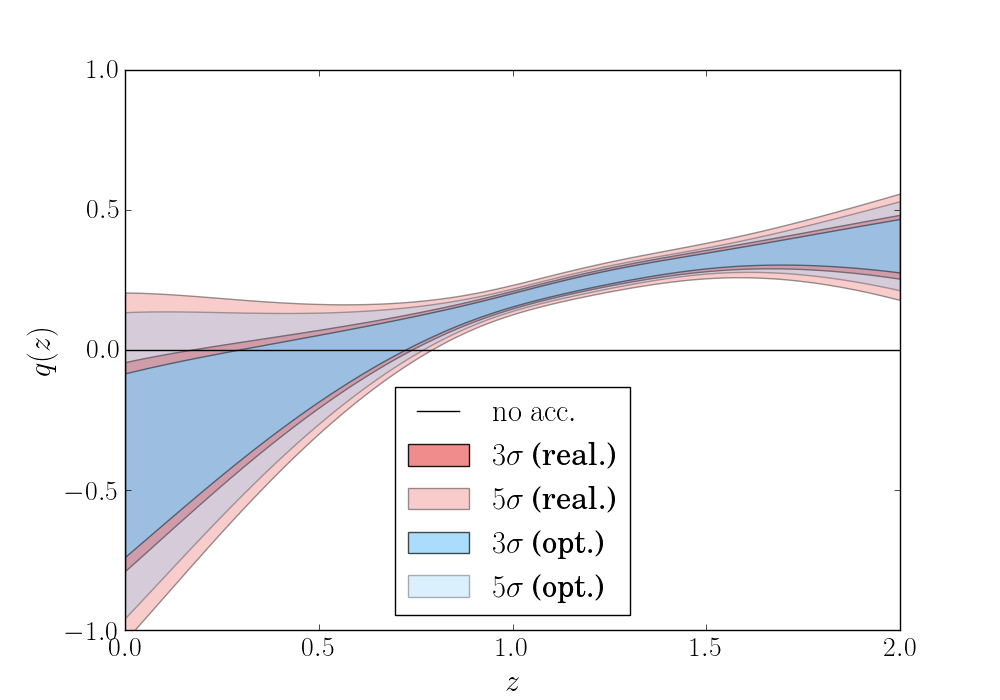}
\caption{Left panel: Gaussian-processes reconstructed $q(z)$ following Eqs.~\eqref{eq:qz} and~\eqref{eq:err_qz} for a SKA-like B1 survey assuming the realistic ($N_1=10$ and $N_2=5$, in blue) and optimistic ($N_1=20$ and $N_2=10$), in red) specifications. The darker (lighter) shaded curves provide the $3\sigma$ ($5\sigma$) confidence levels. The black line denotes shows the non-accelerated threshold at $q_0=0$. Right panel: Same as the left panel, but valid for an Euclid-like survey.} 
\label{fig:rec_qz_nob2}
\end{figure*}

\begin{figure*}
\centering
\includegraphics[width=0.48\textwidth,
height=5.8cm]{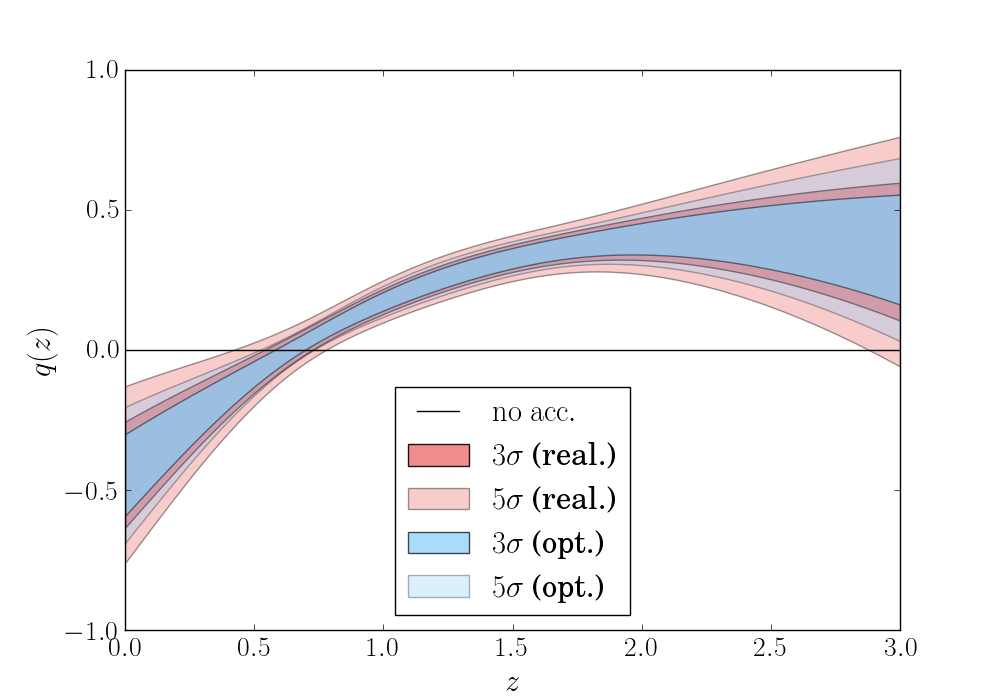}
\includegraphics[width=0.48\textwidth,
height=5.8cm]{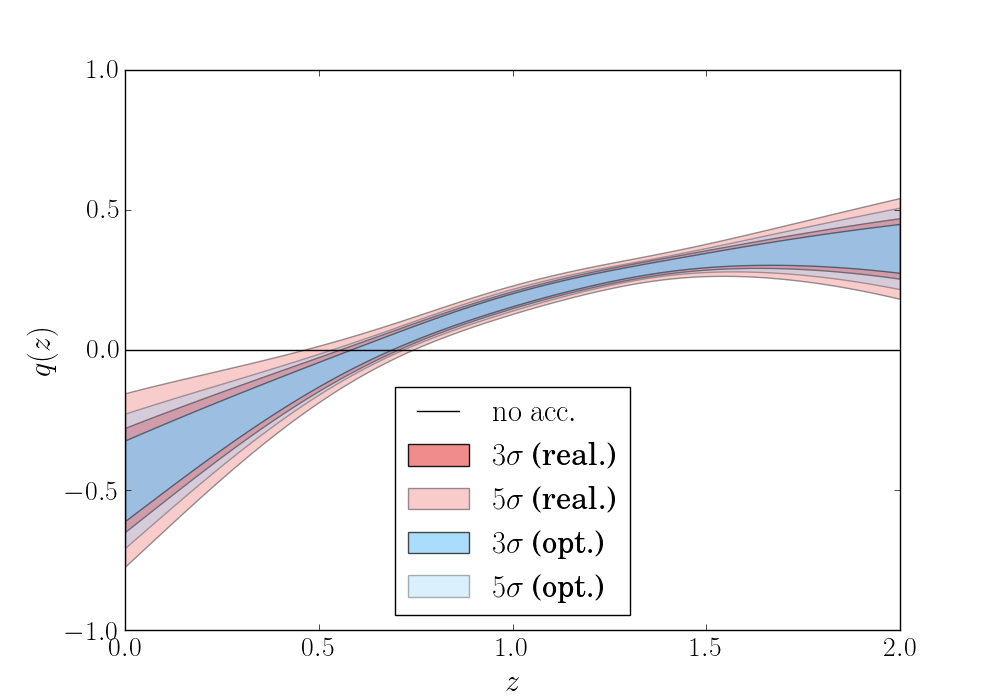}
\caption{Same as Fig.~\ref{fig:rec_qz_nob2}, but including the SKA-like B2 data points.} 
\label{fig:rec_qz_b2}
\end{figure*}

A Gaussian process is a distribution over functions, rather than over variables as in the case of a Gaussian distribution. Therefore, we can reconstruct a function from data points without assuming a parametrisation. We do so using the {\sc GaPP} (Gaussian Processes in Python) code~\citep{Seikel:2012uu} (see also~\citealt{Shafieloo:2012ht}) in order to reconstruct $H(z)$ from data (For other applications of {\sc GaPP} in cosmology, see e.g.~\citealt{Yahya:2013xma, Busti:2014dua, Cai:2015pia, Gonzalez:2017fra, Pinho:2018unz, Gomez-Valent:2018hwc, vonMarttens:2018bvz, Bengaly:2019oxx, Keeley:2019hmw}).  
We simulate $H(z)$ data assuming the fiducial model, 
\begin{eqnarray}\label{eq:hz}
\left[\frac{H(z)}{H_0}\right]^2 &=& \Omega_{\rm m}(1+z)^3 + (1-\Omega_{\rm m}-\Omega_{\rm DE})(1+z)^2 + 
\\ 
&& \Omega_{\rm DE} \exp{\left[3\int_0^z \frac{1+w(z')}{1+z'}dz'\right]}  \;,
\end{eqnarray}
which is valid for a generic dark energy model. We assume the fiducial model to be consistent with Planck 2018 (TT, TE, EE+lowE+lensing) best-fit for flat $\Lambda$CDM, therefore $\Omega_{\rm DE}=\Omega_{\Lambda}=1-\Omega_{\rm m}$, $w(z)=w_0=-1$, and the fiducial values of $H_0$ and the matter density $\Omega_{\rm m}$ are chosen to be
\begin{eqnarray}\label{eq:model1}
H_0 = & 67.36 \pm 0.54 \, \mathrm{km \, s}^{-1} \, \mathrm{Mpc}^{-1} \,, 
\notag\\
\Omega_{\rm m} = & 0.3166 \pm 0.0084 \,. 
\end{eqnarray}
We produce the simulated $H(z)$ measurements by the same fashion of~\cite{Bengaly:2019oxx}, whose specifications for SKA- and Euclid-like surveys follow~\cite{Bacon:2018dui, Amendola:2016saw}:\\
\underline{SKA-like intensity mapping  survey}:
\begin{eqnarray}\label{ska}
\mbox{Band 1:~~} && 0.35<\nu<1.05 \, \mbox{GHz,} \notag\\  
&& 0.35<z<3.06\,, ~ N=10, 15, 20\,, \notag\\
\mbox{Band 2:~~} && 0.95<\nu<1.75 \, \mbox{GHz,} \notag\\  
&& 0.1~<z~<0.5~\,, ~ N=5, 10\,,\notag\\
\mbox{Band 1+2:~~} && N_1=10, 15~\mbox{and}~  N_2=5; \, \notag\\ && N_1=20~\mbox{and}~N_2=10\,,   
\end{eqnarray}
\underline{Euclid-like galaxy survey}:  
\begin{eqnarray} \label{euc}
\mbox{Euclid-like only:~~}  &&  0.6<z<2.0\,, ~ N=10, 15, 20\,, \notag\\
\mbox{Euclid-like + Band 2:~~} && N_1=10, 15~\mbox{and}~ N_2=5; \, \notag\\ && N_1=20~\mbox{and}~N_2=10\,,  
\end{eqnarray}
These prescriptions assume two realistic assumptions for $H(z)$ with Euclid- and SKA-like B1, with $N_1=10$ data points and $N_2=5$ for SKA-like B2, and an optimistic one with $N_1=20$ and $N_2=10$. 
The $H(z)$ measurement uncertainties are taken from the interpolated curves in Figure 10 (left) of~\cite{Bacon:2018dui}.\footnote{The BAO scales which produce the $H(z)$ measurements are within the regime where foreground removal should be very efficient~\citep{Bull:2015lja, Villaescusa-Navarro:2016kbz}}

The deceleration parameter can be obtained from the GP-reconstructed $H(z)$ for each of these survey configurations according to
\begin{equation}\label{eq:qz}
q(z) = -\frac{\ddot{a}}{aH} = (1+z)\frac{H'(z)}{H(z)}-1 \,,
\end{equation}
where $H'(z) \equiv dH(z)/dz$. Its uncertainty is given by error-propagating $q(z)$ with respect to $H(z)$ such as
\begin{equation}\label{eq:err_qz}
\left(\frac{\sigma_{\rm q}}{1+q}\right)^2 =  \left(\frac{\sigma_{H}}{H}\right)^2 + \left(\frac{\sigma_{H'}}{H'}\right)^2 - \left(\frac{2\sigma_{HH'}}{HH'}\right)
\end{equation}
where $\sigma_H$ and $\sigma_{H'}$ represent the uncertainties of the reconstructed $H$ and $H'$, respectively, and $\sigma_{HH'}$ their respective covariance. Conversely from~\cite{Bengaly:2019oxx}, we did not check the results by other future surveys like MeerKat, DESI and SKA galaxy survey because the constraints on $q_0$ would be even more degraded due to the the calculation of $H'$. Same applies for $D(z)$ measurements from the angular mode of BAO, which would involve the second derivative computation of this quantity since $H(z) = 1/D'(z)$. A more thorough assessment of the cosmic acceleration using luminosity distance measurements from forthcoming standard candles and sirens surveys will be pursued in the future.

\section{Results}\label{res}

\begin{table}
    \centering
    \caption{Respectively, the redshift survey, number of data points, the reconstructed $q_0$ value with its respective uncertainty, and how many $\sigma$ away we have $q_0=0$. We can see that SKA-like B1 and Euclid-like surveys can probe $q_0<0$ at over $7\sigma$ level when combined with SKA-like B2. Results obtained with real $H(z)$ measurements are presented as well.}
    \begin{tabular}{ccccccc}
        \hline\hline
        Sample & $N_1$ & $N_2$ & $q_0$ & $\sigma_{q_0}$ & $q_0<0 \; (\sigma)$ \\ \hline\hline
	    Euclid-like & $10$ & - & $-0.416$ & $0.124$ & $3.348$\\
	    & $10$ & $5$ & $-0.464$ & $0.062$ & $7.493$ \\
	    & $15$ & - & $-0.413$ & $0.116$ & $3.568$ \\
	    & $15$ & $5$ & $-0.463$ & $0.059$ & $7.791$ \\
	    & $20$ & - & $-0.411$ & $0.109$ & $3.764$ \\
	    & $20$ & $10$ & $-0.467$ & $0.048$ & $9.736$ \\
	    \hline
	    SKA-like & $10$ & - & $-0.416$ & $0.083$ & $4.952$ \\
	    & $10$ & $5$ & $-0.445$ & $0.063$ & $7.058$ \\
	    & $15$ & - & $-0.413$ & $0.078$ & $5.325$ \\
	    & $15$ & $5$ & $-0.443$ & $0.059$ & $7.486$ \\
	    & $20$ & - & $-0.412$ & $0.073$ & $5.672$ \\
	    & $20$ & $10$ & $-0.447$ & $0.049$ & $9.191$ \\
	    \hline\hline 
	    CC & $30$ & - & $-0.485$ & $0.314$ & $ 1.541$ \\
	    CC+BAO & $30$ & $18$ & $-0.457$ & $0.174$ & $2.631$ & \\
    \end{tabular}
    \label{tab:q0_results}
\end{table}

\subsection{Constraints on $q_0$}

We find that an Euclid-like survey SKA-like B1 survey can probe cosmic acceleration, that is, $q_0<0$, at roughly $3.3-3.8\sigma$ level, whereas a SKA-like B1 survey will be able to do it at $5.0-5.7\sigma$. This is because an Euclid-like survey is expected to cover of a higher redshift range than a SKA-like B1, so the extrapolation to lower redshift ranges worsens. By combining both SKA-like B1 and Euclid-like surveys with SKA-like B2, there is a significant improvement on the $q_0$ constraints due to its lower redshift coverage. For instance, SKA-like B1+B2 combined can determine $q_0<0$ at a $7.0\sigma$ level for a realistic configuration, and $9.2\sigma$ for an optimistic one, and Euclid-like + B2 can do it at a $7.5\sigma$ ($9.7\sigma$) level for a realistic (optimistic) configuration, respectively. 

\begin{figure*}
\centering
\includegraphics[width=0.45\textwidth,
height=5.6cm]{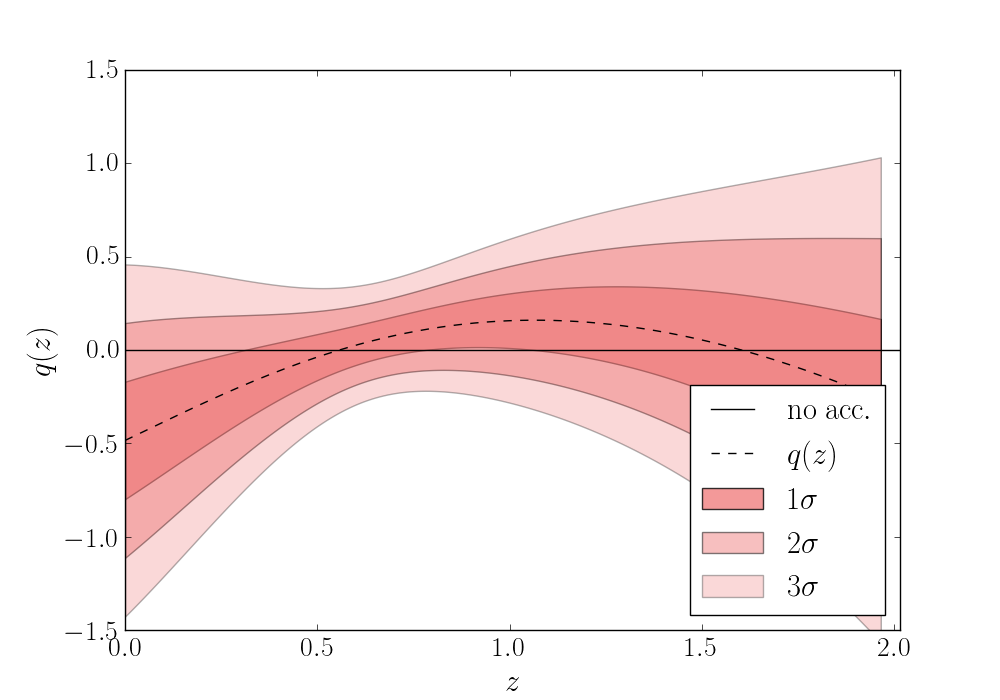}, 
\includegraphics[width=0.45\textwidth,
height=5.6cm]{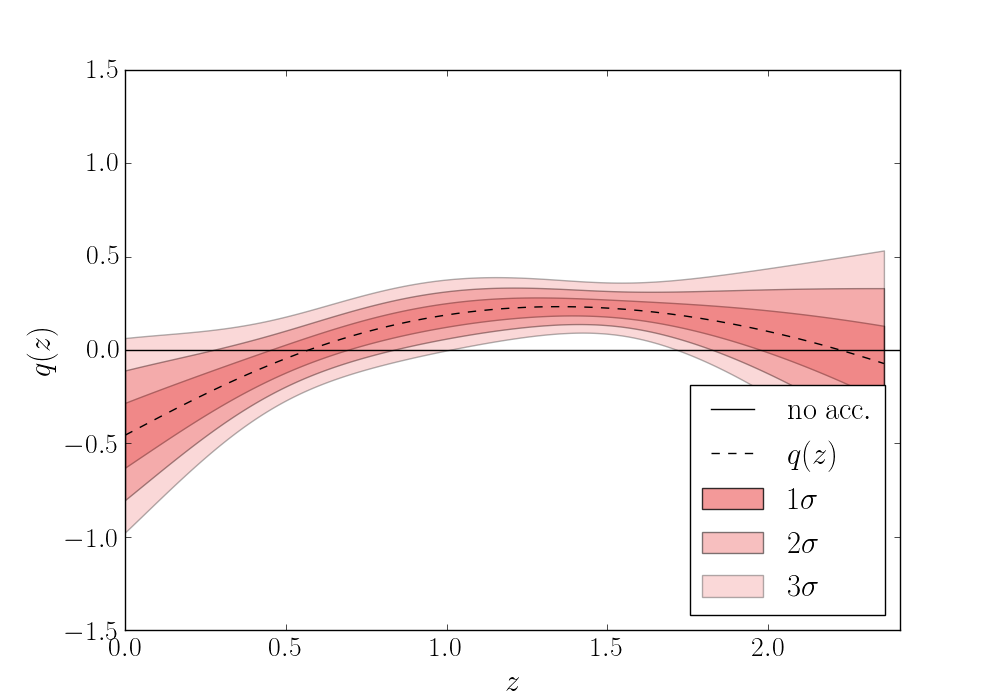}
\caption{The reconstructed $q(z)$ curves, and their $1$, $2$ and $3\sigma$ uncertainties using real $H(z)$ data from CC (left) and CC combined with BAO measurements from galaxy surveys like SDSS and WiggleZ (right).} 
\label{fig:rec_qz_realdata}
\end{figure*}

For the sake of comparison, we perform a $q(z)$ reconstruction obtained with real $H(z)$ data from cosmic chronometres (CC) from differential galaxy ages~\citep{Jimenez:2001gg,Moresco:2016mzx} and from the radial BAO mode of galaxy clustering. We use the $H(z)$ measurements as compiled by~\cite{Magana:2017nfs}, whose results are presented in Table~\ref{tab:q0_results} and Fig.~\ref{fig:rec_qz_realdata}. We obtain that they can only determine if the Universe is currently accelerating at just above $1.5\sigma$ ($2.6\sigma$) level for CC (CC+BAO), and that our results are compatible with previous analysis within $1\sigma$ level~\citep{Haridasu:2018gqm,Arjona:2019fwb}. This demonstrates how Euclid and SKA surveys will largely improve the model-independent assessments of cosmic acceleration evidence. 

\subsection{Robustness tests}

\begin{figure*}
\centering
\includegraphics[width=0.45\textwidth,
height=5.6cm]{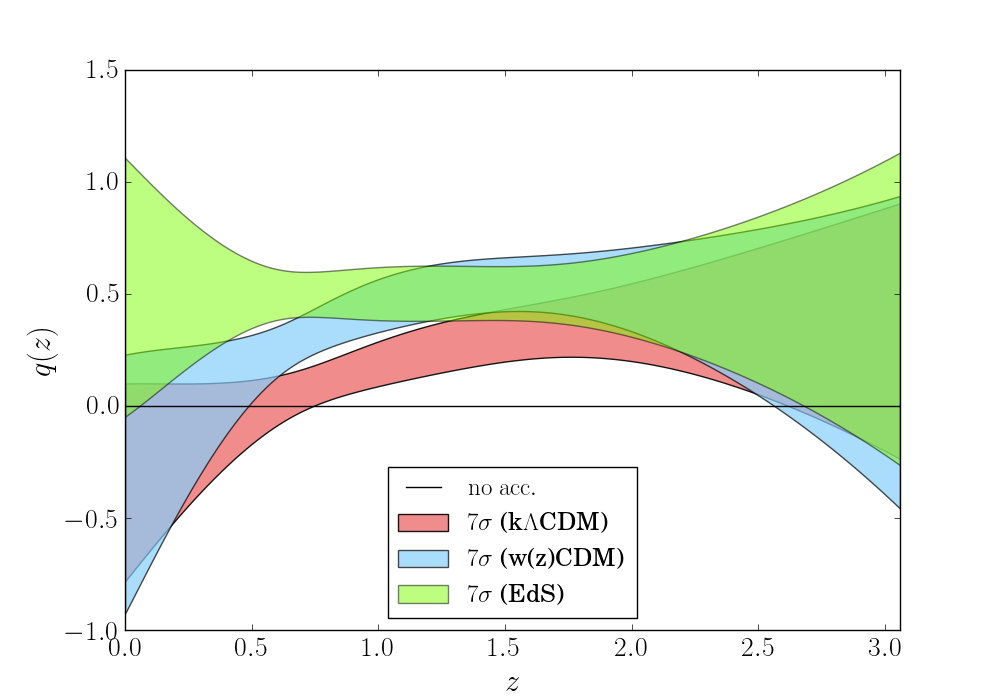}, 
\includegraphics[width=0.45\textwidth,
height=5.6cm]{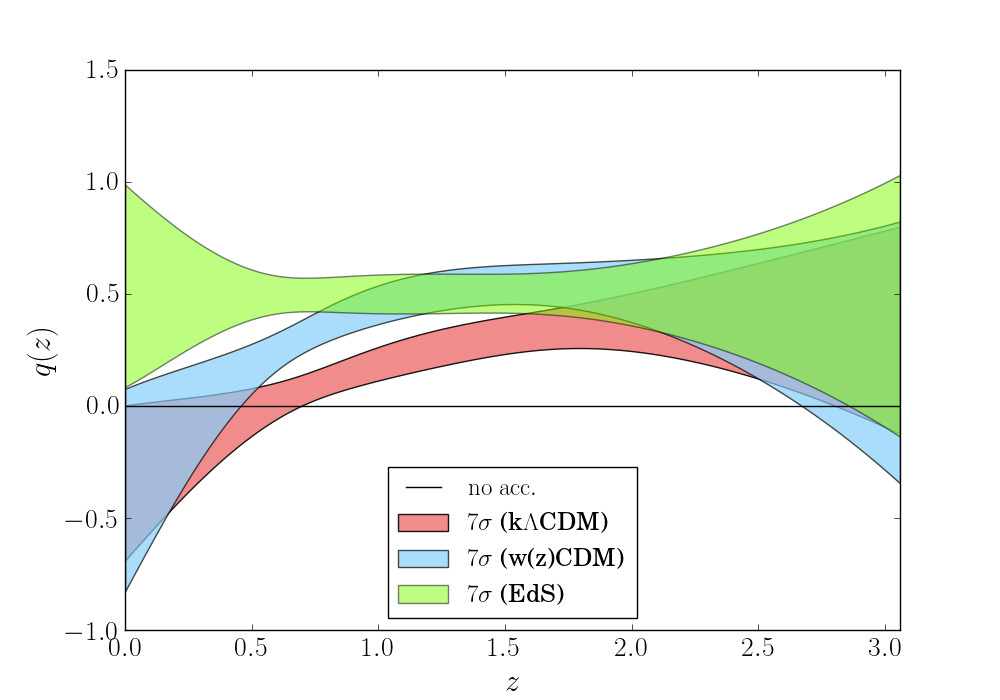}
\caption{The reconstructed $q(z)$ curves (in $7\sigma$) for SKA-like B1 and B2 surveys combined assuming the k$\Lambda$CDM (red), w(z)CDM (blue) and EdS (green) models. The left plot displays the results for a realistic survey specification ($N_1=10$ and $N_2=5$), and the right plot for an optimistic one ($N_1=20$ and $N_2=10$).} 
\label{fig:rec_qz_othermodels}
\end{figure*}

We test how robust are our results with respect to other cosmological models than flat $\Lambda$CDM. We thus produce data-sets for the optimistic specifications of SKA-like B1 and Euclid-like surveys combined with the SKA-like B2 assuming the following models:
\begin{eqnarray} \label{models}
\mbox{w(z)CDM:~~} & &  w(z) = (-1/2) + (1/2)\tanh{[3(z-1/2)]}\,,
\notag\\
\mbox{k$\Lambda$CDM:~~} & & \Omega_{\rm k} \equiv 1-\Omega_{\rm m}-\Omega_{\rm DE}=-0.10 \,,
\notag\\
\mbox{EdS:~~} & & \Omega_{\rm m}=1,\; \Omega_{\rm DE}=0 \,,
\end{eqnarray}
where the last model, EdS, stands for Einstein-de Sitter model that gives $q_0=0.5$, i.e., a non-accelerated model consistent with the cosmic expansion at matter-dominated era. We use it for the sake of determining how well can we rule it out from the standard model given the precision of this data. 

The $q_0$ constraints obtained for these cases are all consistent with the fiducial model, and with relative uncertainties compatible with the standard model analysis. For a SKA-like B1 + B2 survey with optimistic (realistic) configurations, we obtained $q^{\mathrm{w(z)CDM}}_0=-0.375 \pm 0.065$ ($q^{\mathrm{w(z)CDM}}_0=-0.349 \pm 0.082$), $q^{\mathrm{k}\Lambda\mathrm{CDM}}_0=-0.344 \pm 0.049$ ($q^{\mathrm{k}\Lambda\mathrm{CDM}}_0=-0.341 \pm 0.063$), and $q^{\mathrm{EdS}}_0=+0.536 \pm 0.072$ ($q^{\mathrm{EdS}}_0=+0.530 \pm 0.092$). We present these results in Fig.~\ref{fig:rec_qz_othermodels}. Hence, we show that we will be able to rule out a non-accelerated model like EdS at over $11\sigma$ ($8\sigma$) level with a SKA-like survey using $H(z)$ data {\it alone}. Similar results were obtained for an Euclid-like survey.

We checked if the assumption of a fixed fiducial cosmological model affects our results. We produced Monte Carlo realisations varying the cosmological parameters $\bm{p}=(\Omega_{\rm m},H_0)$ according to a Gaussian distribution $\mathcal{N}(\bm{p}, \sigma_{\bm{p}})$, where the parameters and their uncertainties are given by Eq.~\ref{eq:model1}. We found that the measured $q_0$ values are fully compatible with the uncertainties quoted in Table~\ref{tab:q0_results}. Finally, we verified how our results change with respect to other GP kernels. For a SKA-like B1+B2 survey assuming a realistic ($N_1=10$ and $N_2=5$) configuration, for example, we obtained $q_0=-0.433 \pm 0.076$ and $q_0=-0.437 \pm 0.070$ for a Mat\'ern(7/2) and Mat\'ern(9/2) kernel, respectively, thus the detection of current cosmic acceleration would be at a $5.6\sigma$ $(6.2\sigma)$ level. As for the optimistic survey specification, again assuming these respective kernels, we found a $7.1\sigma$ $(8.0\sigma)$ evidence positive acceleration instead. So we will still be able to probe cosmic acceleration at over $5\sigma$ level even for the most conservative kernels.


\section{Discussion and concluding remarks}\label{disc}

The nature of the late-time accelerated expansion of the Universe remains one of the most intriguing phenomenon. Although the Cosmological Constant can account for that and explain the observations with unprecedented precision thus far, it suffers from many theoretical problems. Most of viable alternatives to the Cosmological Constant are also plagued with similar issues. In addition, there are still discussions whether the Universe is truly undergoing an accelerated expansion today depending on how one approaches the available data~\citep{Nielsen:2015pga, Colin:2018ghy}. Nonetheless, all this debate relies on fitting cosmological models with observations. It is essential to quantify the evidence for cosmic acceleration in a model-independent way - it will not tell us the {\it best} model that describes observations, but it will be able to underpin (or rule out) this phenomenon regardless of the underlying model.  

We simulated data for next-generation redshift surveys for this purpose. We produced synthetic $H(z)$ measurements reproducing SKA- and Euclid-like radial BAO measurements, and hence perform a non-parametric reconstruction of the $H(z)$ and $H'(z)$ for the sake of providing $q(z)$, the deceleration parameters - so that its current value, $q_0$, will tell if the Universe is currently accelerating or not. We found a $\sim 5\sigma$ evidence for this result with a SKA-like B1 survey, and $\sim 3\sigma$ for an Euclid-like one. When combining both experiments with the SKA-like B2 survey, which will probe a lower redshift threshold, the evidence increases to at least $7\sigma$ level. 
These results are consistent with currently available observations, but {\it without} the implicit assumption of $\Lambda$CDM or any other dark energy model. 

These results demonstrate the capability of next-generation redshift surveys on underpinning the evidence for cosmic acceleration in a truly model-independent way, and as powerful probes of late-time Cosmology for the future, along with standard candles and sirens.  

\subsection*{Acknowledgements}
CB thanks Roy Maartens, Chris Clarkson, Jailson S. Alcaniz, and Martin Kunz for valuable comments and discussions. CB acknowledges support from the  South African Radio Astronomy Observatory
(SARAO) at an early stage of this work, as well as from the Swiss National Science Foundation. 


\bsp	
\label{lastpage}
\end{document}